\documentclass{article}

\PassOptionsToPackage{numbers, compress}{natbib}
\usepackage[final]{tackling_climate_workshop_style}

\usepackage[utf8]{inputenc} % allow utf-8 input
\usepackage[T1]{fontenc}    % use 8-bit T1 fonts
\usepackage{hyperref}       % hyperlinks
\usepackage{url}            % simple URL typesetting
\usepackage{booktabs}       % professional-quality tables
\usepackage{amsfonts}       % blackboard math symbols
\usepackage{nicefrac}       % compact symbols for 1/2, etc.
\usepackage{microtype}      % microtypography
\usepackage{amsmath}
\usepackage{mathtools}
\usepackage{wrapfig}
\usepackage{xcolor}

\title{Southern Ocean Dynamics Under Climate Change: New Knowledge Through Physics-Guided\\ Machine Learning}

\author{%
  William Yik \\
  Dept. of Computer Science/Mathematics \\
  Harvey Mudd College\\
  Claremont, USA \\
  \texttt{wyik@hmc.edu} \\
  \And
  Maike Sonnewald \\
  Dept. of Computer Science \\
  University of California \\
  Davis, USA \\
  \texttt{sonnewald@ucdavis.edu} \\
  \AND
  Mariana C. A. Clare \\
  ECMWF \\
  Bonn, Germany \\
  \texttt{mariana.clare@ecmwf.int} \\
  \And
  Redouane Lguensat \\
  IPSL, IRD \\
  Paris, France \\
  \texttt{rlguensat@ipsl.fr} \\
}

\begin{document}

\maketitle

\begin{abstract}
Complex ocean systems such as the Antarctic Circumpolar Current play key roles in the climate, and current models predict shifts in their strength and area under climate change. However, the physical processes underlying these changes are not well understood, in part due to the difficulty of characterizing and tracking changes in ocean physics in complex models. Using the Antarctic Circumpolar Current as a case study, we extend the method Tracking global Heating with Ocean Regimes (THOR) to a mesoscale eddy permitting climate model and identify regions of the ocean characterized by similar physics, called dynamical regimes, using readily accessible fields from climate models. To this end, we cluster grid cells into dynamical regimes and train an ensemble of neural networks, allowing uncertainty quantification, to predict these regimes and track them under climate change. Finally, we leverage this new knowledge to elucidate the dynamical drivers of the identified regime shifts as noted by the neural network using the `explainability' methods SHAP and Layer-wise Relevance Propagation. A region undergoing a profound shift is where the Antarctic Circumpolar Current intersects the Pacific-Antarctic Ridge, an area important for carbon draw-down and fisheries. In this region, THOR specifically reveals a shift in dynamical regime under climate change driven by changes in wind stress and interactions with bathymetry. Using this knowledge to guide further exploration, we find that as the Antarctic Circumpolar Current shifts north under intensifying wind stress, the dominant dynamical role of bathymetry weakens and the flow intensifies.
\end{abstract}

\section{Introduction}\label{sec: intro}
Complex ocean systems such as the Antarctic Circumpolar Current (ACC) play a central role in the climate through mechanisms such as heat transport via powerful currents, large-scale overturning, and carbon exchange with the atmosphere \cite{chapman2020defining,morrison2013relationship,boning2008response}. Such systems are known to exhibit a wide array of responses to anthropogenic forcing, but such changes and their underlying physics are poorly constrained \cite{boning2008response, langlais2015sensitivity, larson2020extracting}. This is, in part, a direct consequence of difficulties in both characterizing ocean physics and tracking its shifts under climate change. Model intercomparison projects, such as the Coupled Model Intercomparison Project (CMIP6) \cite{meehl2000coupled,eyring2016overview,o2016scenario}, have set up a vital framework for understanding differences between climate models in both of these areas.  A key target for the sixth and current phase of CMIP was to increase the horizontal resolution, which in the ocean partially allows for mesoscale turbulence to be expressed. However, modern models have only exacerbated challenges in characterizing and tracking ocean physics due to their complexity and size. As a result, key systems such as the ACC as well as their variability among models are often described using bulk metrics, such as, the overturning strength integrated over global longitude \cite{chapman2020defining,donohue2016mean}. Bulk metrics have limited scope for elucidating the dynamics driving the changes, as these largely vary in space. For the first time, we apply the method Tracking global Heating with Ocean Regimes (THOR) proposed by \cite{sonnewald2021revealing} to the ocean of a mesoscale turbulence-permitting model, the Coupled Model 4 \cite{held2019structure,adcroft2019gfdl}, to both understand and track ocean dynamics under climate change. THOR is a three-step process beginning with unsupervised classification of ocean grid cells into dynamical regimes based on their physics. For this step, we use Native Emergent Manifold Interrogation (NEMI) \cite{sonnewald2023hierarchical}. The second component of THOR trains a deep ensemble of neural networks to predict these dynamical regimes from the previous step using readily accessible fields as inputs (e.g., sea surface height, depth, wind stress, and mass transport). Finally, the third step of THOR applies this trained deep ensemble to a new climate change scenario or entirely new model of interest to understand changes in dynamical regimes over time. This step also incorporates uncertainty quantification and eXplainable Artificial Intelligence (XAI) methods to guide further exploration of newly discovered ocean phenomena. Here we use THOR to investigate changes in the the region where the ACC meets the Pacific-Antarctic Ridge (PAR). THOR specifically reveals a shift in dynamical regime in this region under climate change, and XAI methods additionally show that these are related to changes in wind stress and interactions with the bathymetry. Using this new information provided by THOR, we further explore the ACC under climate change and find that the shifts in regional dynamical regime is explained by northward movement of the ACC driven by changes in wind stress. This ACC movement brings it into a new, less variable bathymetric region where its interactions with the sea floor are less strong, thus leading to a flow distributed over a larger area and concentrated at the surface. The remainder of this manuscript is structured as follows. Section \ref{sec: THOR} provides a more detailed overview of the THOR method. Section \ref{sec: application} details our exploration of the ACC guided by new knowledge revealed by THOR. Lastly, Section \ref{sec: conclusion} concludes the paper with potential directions for future work.

\section{Tracking global Heating with Ocean Regimes (THOR)}\label{sec: THOR}
\begin{wrapfigure}[16]{r}{0.5\textwidth}
    \vspace{-0.75in}
    \centering
    \includegraphics[width=0.5\textwidth]{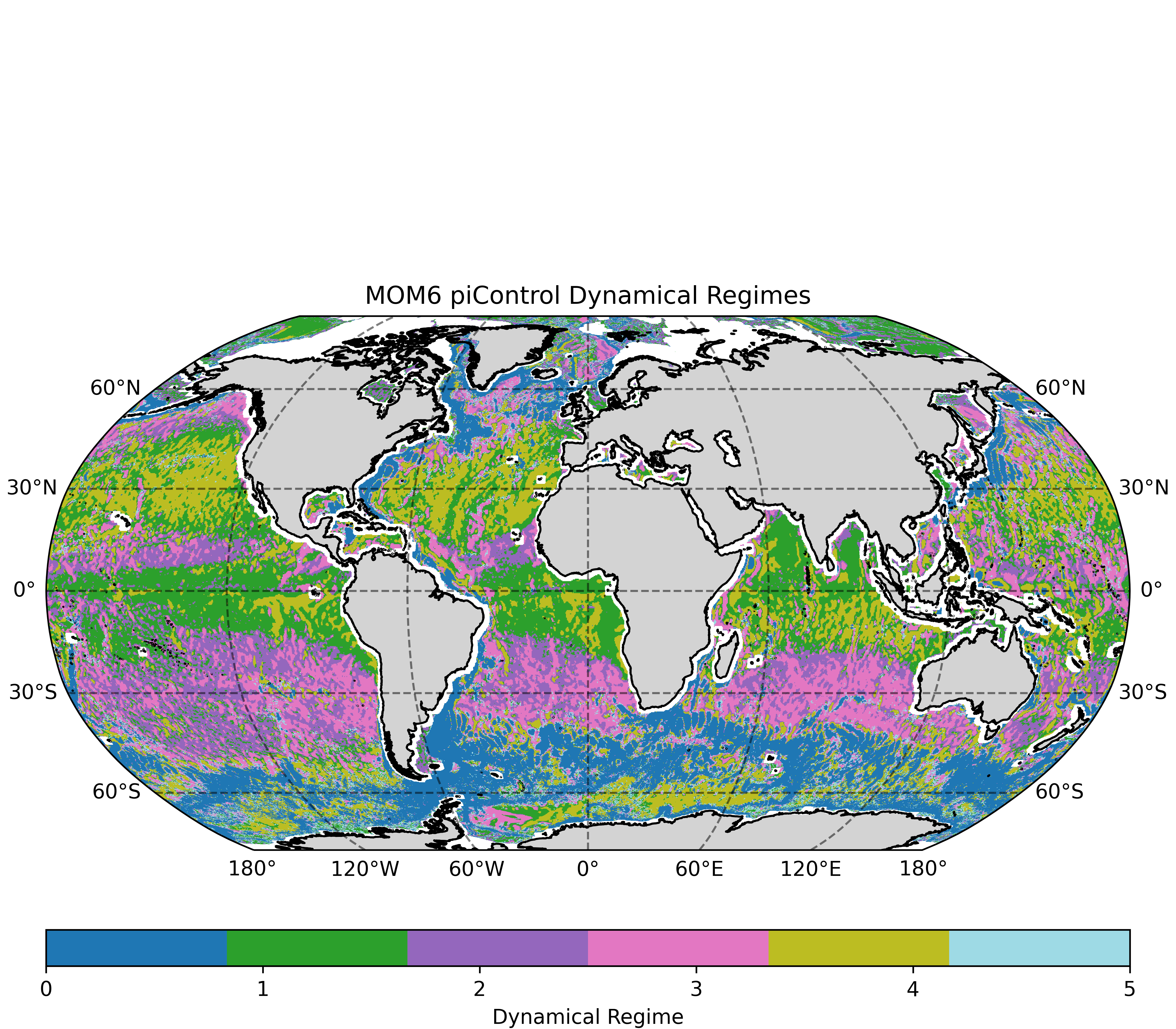}
    \caption{Six Dynamical regimes discovered by NEMI on the piControl run of MOM6.}
    \label{fig: mom6 piControl regimes}
\end{wrapfigure}

\begin{figure}[ht]
    \centering
    \includegraphics[width=\textwidth]{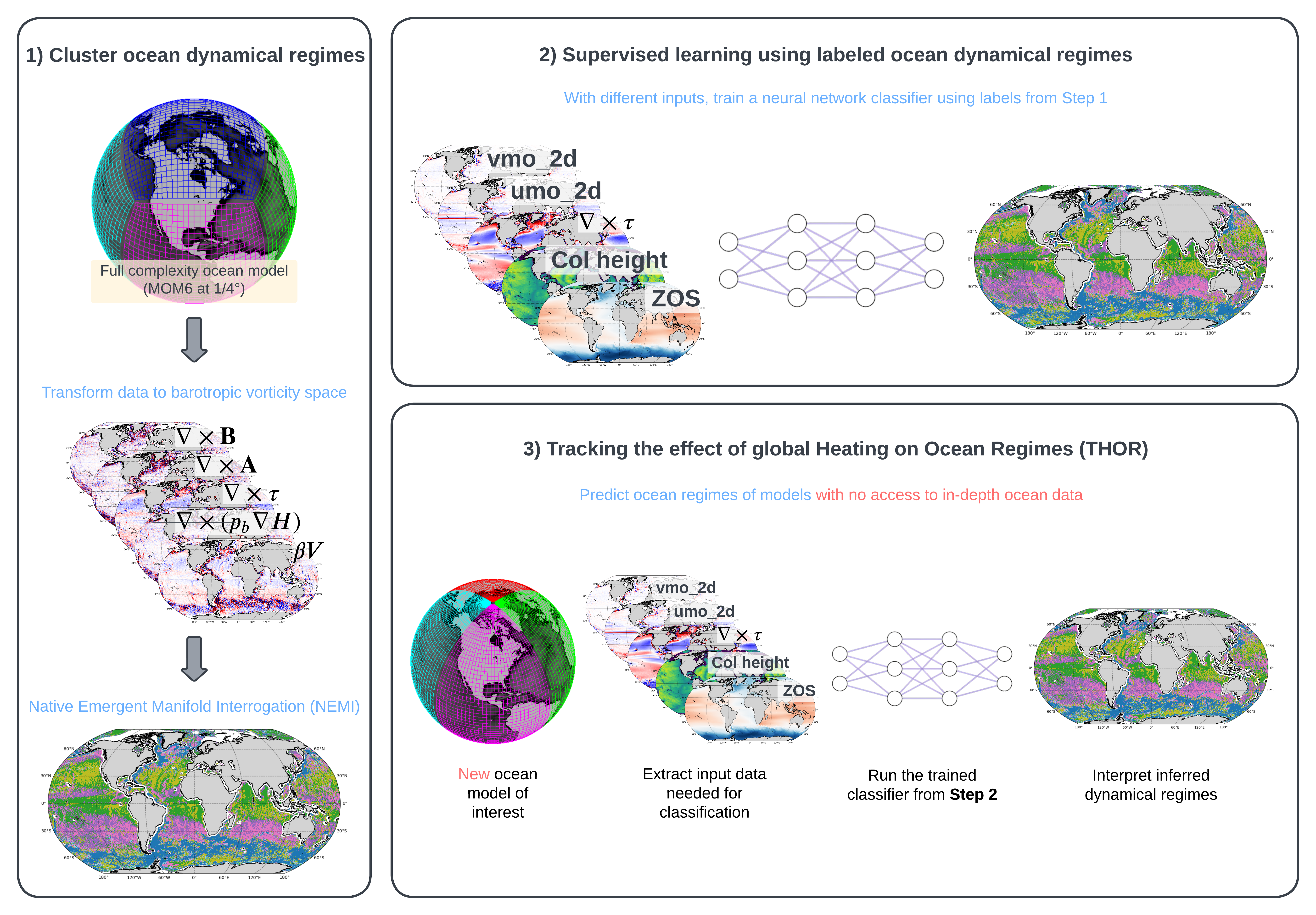}
    \caption{A high-level overview of the THOR workflow for identifying and tracking ocean dynamical regimes. Figure adapted from \cite{sonnewald2021revealing}.}
    \label{fig: thor diagram}
\end{figure}

In this manuscript we apply THOR (Figure \ref{fig: thor diagram}) to the Modular Ocean Model version 6 (MOM6) \cite{griffies2012elements,adcroft2019gfdl,griffies2020primer}, a component of the Coupled Model version 4 (CM4) \cite{held2019structure}. CM4 operates at an ocean resolution of $0.25^\circ$ which allows for mesoscale eddies, differentiating itself from the $1^\circ$ ECCO model \cite{forget2015ecco} studied by \cite{sonnewald2021revealing}. The first step of THOR addresses the key issue of unlabeled data in ML applications for climate sciences by leveraging unsupervised learning to create a labeled dataset from high dimensional ocean model data. Specifically, ocean grid cells are clustered into groups, called dynamical regimes, based on their mean balance of the barotropic vorticity (BV) equation, a characterization of local circulation of fluid flow, throughout a preindustrial control (piControl) run, illustrating the state of the ocean before the start of anthropogenically induced climate change. For more details, see \cite{sonnewald2021revealing,khatri2023scale}.

Previous applications of THOR to the ECCO model \cite{sonnewald2021revealing,sonnewald2021icml} used k-means clustering to reveal dynamical regimes. However, \cite{sonnewald2023hierarchical} demonstrates that k-means fails to converge for several information criteria metrics on the higher resolution MOM6 data. As such, for our first step of THOR (Figure \ref{fig: thor diagram}a) we substitute k-means for Native Emergent Manifold Interrogation (NEMI), which has been shown by \cite{sonnewald2023hierarchical} to successfully cluster MOM6 BV data into a user-defined number of interpretable dynamical regimes. Here, we choose six labels, or dynamical regimes, following \cite{sonnewald2021revealing}. The global regimes found by NEMI for the piControl run of CM4 are shown in Figure \ref{fig: mom6 piControl regimes}. Notice the oceanographic features which are reflected in the dynamical regimes such as large, wind stress-driven gyres and the the mid-Atlantic ridge. Additionally, note that the vast majority of grid cells are clustered into Regimes 0-4. This indicating that these capture the dominant dynamical regimes in the ocean \cite{sonnewald2023hierarchical} and that Regime 5 represents residual nonlinear regimes, as in \cite{sonnewald2019unsupervised}.

While the first step of THOR is useful for revealing areas of the ocean characterized by similar physics, it requires that the BV budget be closed for each new ocean model and experiment in order for NEMI clustering to be done. This is infeasible for a wide array of models and impossible with the data routinely made available through CMIP. As such, the second step of THOR (Figure \ref{fig: thor diagram}b) trains a neural network (NN) to predict the clustered dynamical regimes from above from more accessible fields from an ocean model. Specifically, the inputs for each grid cell are the sea surface height above the geoid (ZOS), its $x$- and $y$-gradients, the depth relative to sea level (bathymetry), its $x$- and $y$-gradients, the curl of surface wind stress torque ($\nabla \times \tau_s$), the Coriolis parameter ($f$), and the depth-summed zonal and meridional mass transport (umo\_2d and vmo\_2d). The notable additions to this dataset (differentiating it from that of \cite{sonnewald2021revealing}) are the mass transport fields. These were added after preliminary testing showed that neural networks were unable to accurately classify pairs of regimes whose dominant physics were correlated with mass transport. Each of these inputs was chosen because they directly impact the values of BV equation terms, meaning that there is some physical relation between the chosen inputs and the dynamical regimes found in the first step of THOR. Thus, eXplainable artificial intelligence (XAI) methods may be applied to attribute NN predictions to specific physical inputs, allowing us to gain insight into the drivers of regime shift as demonstrated in the next section. During initial testing XAI also revealed spurious correlations between the input fields and output regimes which guided our choice to add the mass transport variables to the input.

We use the same MLP architecture described in \cite{sonnewald2021revealing} to predict the dynamical regimes, and an ensemble of 50 such MLPs are trained in order to regularize the latent loss space and gauge uncertainty in regime predictions. Specifically, we use entropy across the ensemble's predictions for each grid cell to quantify uncertainty following \cite{clare2022explainable}. More detailed results from training can be found in Section \ref{appendix: nn training}. Importantly, we find that though the NN does not achieve exceptionally high accuracy, the average entropy of incorrectly classified grid cells is higher than that of correctly classified grid cells. This indicates that the NN is making well-calibrated predictions, correctly represents its own uncertainty, and is not erroneously overconfident in its predictions.

\section{Application to the Southern Ocean}\label{sec: application}
\begin{figure}[ht]
    \vspace{-0.25in}
    \centering
    \includegraphics[width=\textwidth]{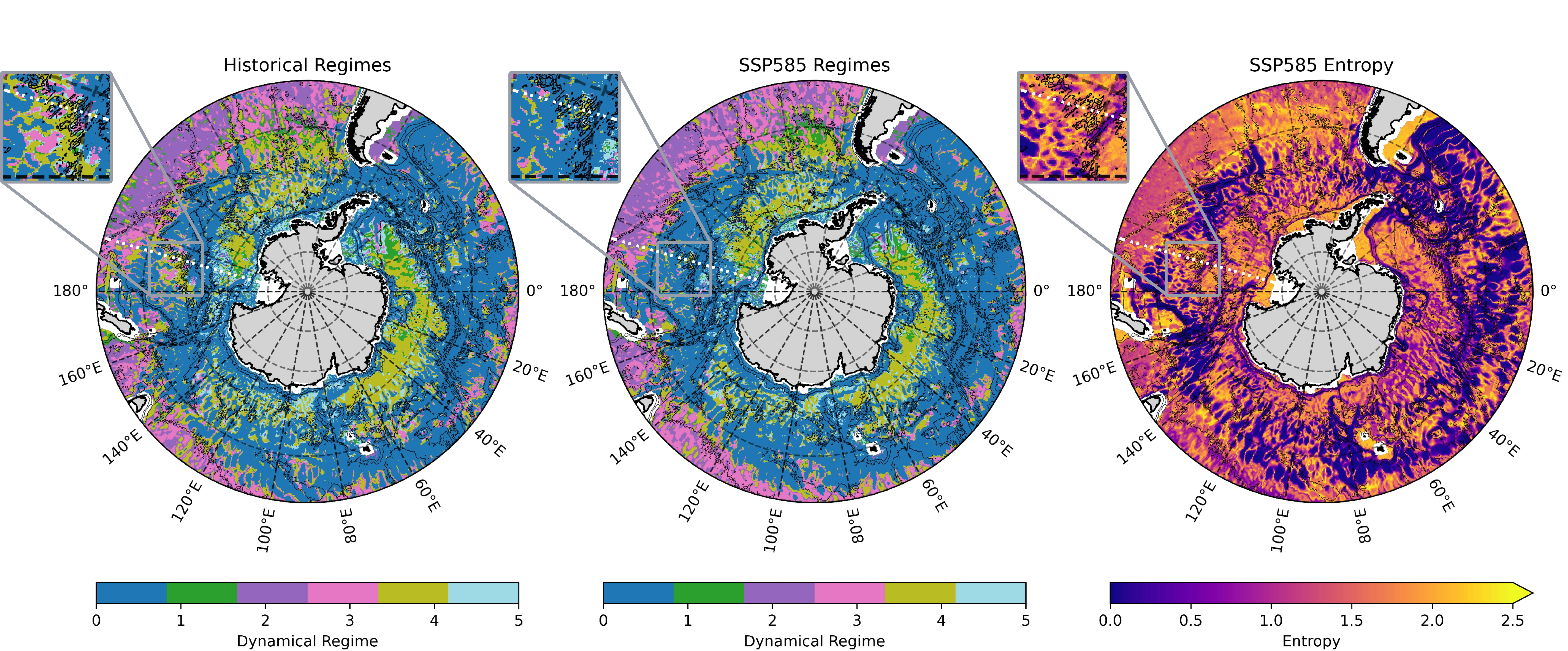}
    \caption{NN dynamical regime predictions for the Historical and SSP585 scenarios along with entropy (uncertainty) for the SS585 predictions. The contours show bathymetry. The inset shows the area of interest where the ACC meets the PAR, and the dashed white line shows the meridian where the transects of Figure \ref{fig: umo transects} are taken.}
    \label{fig: nn historical and ssp585}
\end{figure}

\begin{wrapfigure}[32]{r}{0.5\textwidth}
    \vspace{-0.13in}
    \centering
    \includegraphics[width=0.5\textwidth]{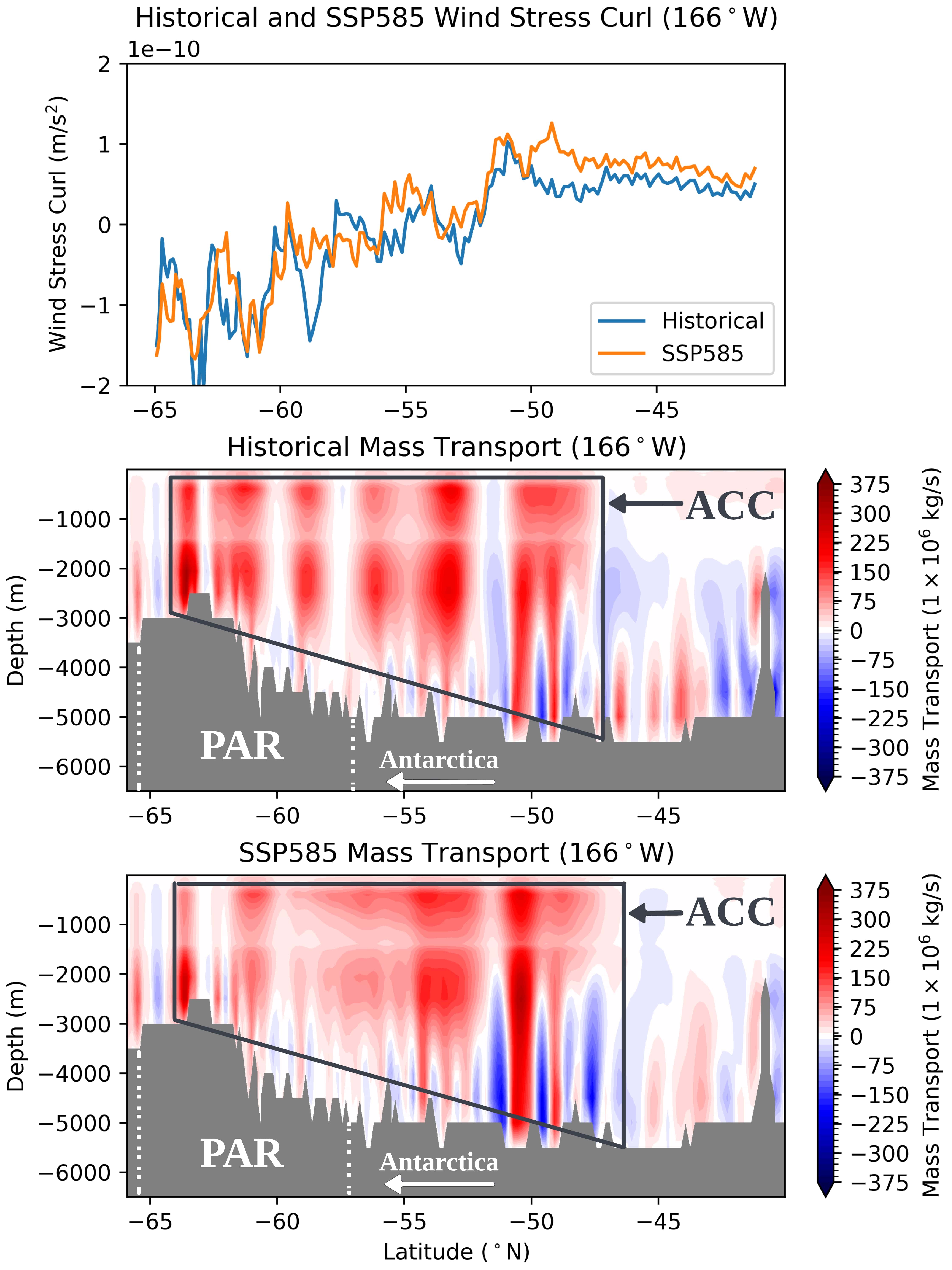}
    \caption{Wind stress curl and lateral (east-west) mass transport for both the Historical and SSP585 scenarios. The ACC and PAR are outlined in black and white, respectively. Red and blue indicate eastward and westward flow, respectively. Note that there is an artifact around -1500m due to regridding.}
    \label{fig: umo transects}
\end{wrapfigure}

The third and final step of THOR (Figure \ref{fig: thor diagram}c) applies the trained NN ensemble from the previous section to predict dynamical regimes for entirely new experiments or ocean models. Here, we apply the ensemble to the Historical (post industrial revolution to 2015) and SSP585 (aggressive anthropogenic warming post 2015) runs of CM4 in order to track regime shifts in the Southern Ocean under climate change. We analyze the 1992-2013 window of the Historical scenario following \cite{sonnewald2021revealing} and the 2080-2100 window of the SSP585 scenario to elucidate the effects of anthropogenic warming. Figure \ref{fig: nn historical and ssp585} shows the deep ensemble's mean Southern Ocean regime predictions for the Historical and SSP585 scenarios as well as the entropy for the SSP585 predictions as a proxy for the ensemble's uncertainty in its predictions under climate change. Acting as a tool to guide new discovery, THOR reveals a regime change where the ACC meets the Pacific-Antarctic Ridge (PAR), a divergent tectonic plate boundary characterized by rough bathymetry at around $60^\circ$S, $166^\circ$W. This region of interest is shown in the insets of Figure \ref{fig: nn historical and ssp585}. The notable shift revealed by THOR from Regime 4 (light green) to Regime 0 (dark blue), combined with the high entropy motivates further investigation.

We use two XAI methods, the NN specific layer-wise relevance propagation (LRP) \cite{montavon2019layer} and model agnostic Shapley Additive exPlanations (SHAP) \cite{garcia2020shapley}, to determine which inputs help THOR's deep ensemble make its predictions where the ACC meets the PAR. The two values generally agreed on the relevance of the surface wind stress curl and the bathymetry, as well the sea surface height which is closely related to the wind (see Section \ref{appendix: xai}). Given that the ACC is largely driven by westerly winds surrounding Antarctica, the XAI methods may be indicating a shift in the wind stress and the interactions between the ACC and the bathymetry. Motivated by this, we investigate lateral (east-west) mass transport at $166^\circ$W, which represents the portion of the ACC which cuts through the PAR, as shown in the bottom two panels Figure \ref{fig: umo transects}. Here, we see a clear shift in the core of the ACC northward away from the PAR. Specifically, in the Historical scenario the eastward jets are closely connected to the bathymetric features of the PAR and concentrated south of diffuse circulation without a jet-like structure (intense red) at $53^\circ$S. In the SSP585 scenario, however, the area of intense flow around $50^\circ$S strengthens markedly and is associated with a strengthening of the wind stress. As such, the core of the ACC moves into a region of less dynamic bathymetry, allowing it to flow more freely and increase in intensity. Aligning with the new knowledge of wind stress relevance, the local wind stress curl strengthens just north of the PAR as shown in the top panel of Figure \ref{fig: umo transects}. In response, the ACC shifts in the same direction, pulling it away from the bathymetric influence of the PAR and fundamentally changing its governing physics, as reflected in the regime shift initially revealed by THOR in Figure \ref{fig: nn historical and ssp585}.

\vspace{-0.1in}
\section{Conclusion}\label{sec: conclusion}
We extend the machine learning method THOR to a mesoscale eddy-permitting climate model in order to gain insight into the drivers underlying changes in ocean physics under climate change. Importantly, we use THOR not as an oracle, but rather a tool to discover new knowledge about ocean dynamical regimes and guide further investigation of shifts in the driving forces behind ocean circulation. Under this framework, THOR reveals that a fundamental shift in Southern Ocean physics occurs under climate change where the ACC meets the PAR. Using this new knowledge combined with XAI methods as a guide, we find that the wind stress curl increases in strength north of the PAR, pulling the ACC with it and effectively moving the ACC from a bathymetrically locked state over the PAR to one dominated by the wind stress where its flow increases in strength. It is important to note that while such conclusions may theoretically be reached by analyzing the raw ocean model data, such analysis is largely infeasible due to the high resolution and dimensionality of the model output. THOR addresses these key issues and enables us to easily identify areas of interest for further investigation such as those discussed in Section \ref{sec: application}.

There are several possible directions for future work. First, different CMIP models have different representations of ocean physics, and the high-resolution version of THOR presented in this work could provide insight into how these differences affect predictions for various ocean systems under climate change. Such insight is at the heart of lowering model spread and uncertainty of future projections. Additionally, since NEMI can cluster the MOM6 data into any given number of regimes, it is possible to train the NNs of THOR to identify more than 6 dynamical regimes discussed in Section \ref{sec: THOR}. Future work will investigate additional oceanographic insights revealed by THOR when analyzing a greater number of regimes. Lastly, the NN ensemble applied in this work classifies the dynamical regimes of grid cells based on local information to that grid cell. Spatially-aware NN architectures may be applied, but the large continents and difficulties applying XAI methods to more complex architectures pose exciting future challenges. Regardless of the supervised technique used to classify dynamical regimes, we maintain that its predictive power should not trade off with its interpretability and transparency.

\vspace{-0.1in}
\begin{ack}
The authors would like to thank Stephen Griffies, Tarun Verma, and Isaac Held for their helpful discussions and feedback. WY acknowledges funding from the NOAA Ernest F. Hollings Undergraduate Scholarship. MS funding: Cooperative Institute for Modeling the Earth System, Princeton University, under Award NA18OAR4320123 from the National Oceanic and Atmospheric Administration, U.S. Department of Commerce.
\end{ack}

\section*{Data Availability Statement}
The data and code are publicly available at \url{https://github.com/yikwill/THOR-MOM6}.

\small
\setcitestyle{numbers}
\bibliographystyle{unsrtnat}
\bibliography{references}
\normalsize

\newpage
\appendix
\section{Supplemental Material}
\subsection{Neural Network Training Details}\label{appendix: nn training}
For the second step of THOR, we train an ensemble of 50 MLP NNs each with the same architecture as \cite{sonnewald2021revealing} but different weight and bias initializations. These NNs are trained to predict dynamical regimes for individual ocean grid cells from accessible ocean model fields. Specifically, the 10 inputs are the sea surface height above the geoid (ZOS), its $x$- and $y$-gradients, the depth relative to sea level (bathymetry), its $x$- and $y$-gradients, the curl of surface wind stress torque ($\nabla \times \tau_s$), the Coriolis parameter ($f$), and the depth-summed zonal and meridional mass transport (umo\_2d and vmo\_2d). The input size of the NNs was resized from 8 in \cite{sonnewald2021revealing} to 10 to account for our addition of the mass transport fields. The outputs are the dynamical regime classifications given by NEMI in step 1 of THOR.

\begin{figure}[ht]
    \centering
    \includegraphics[width=\textwidth]{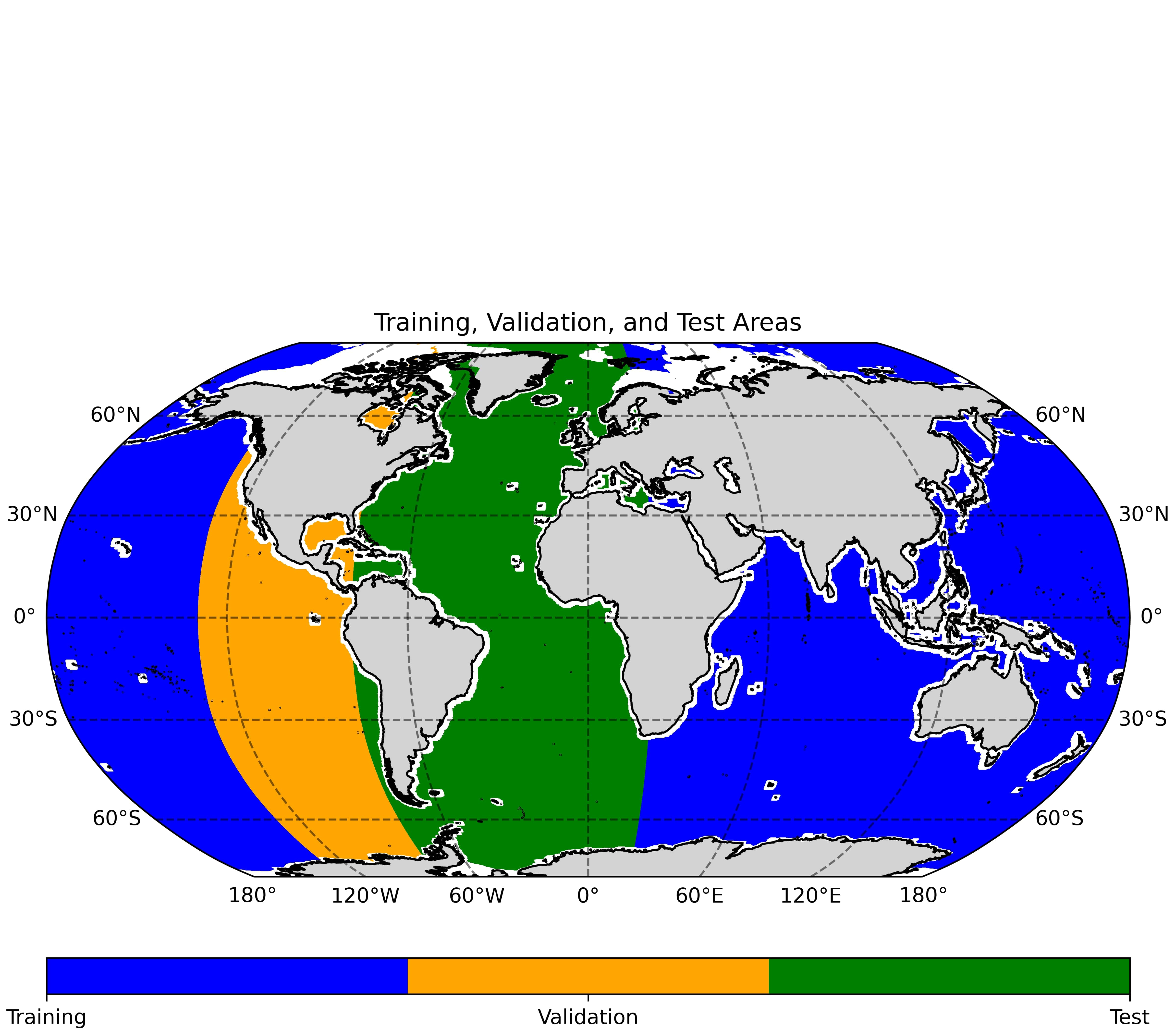}
    \caption{The training, validation, and test areas used for THOR's deep ensemble.}
    \label{fig: train val test areas}
\end{figure}

We train THOR's deep ensemble to predict the dynamical regimes of single grid cells in a preindustial control (piControl run), which represents the state of the ocean before anthropogenic climate change. Thus, the inputs for each grid cell are the piControl mean values of the 10 inputs described above. Similarly, the outputs are one of six dynamical regimes found by NEMI's clustering, which was based on the mean balance of the BV equation during the piControl run. We split the training, validation, and test data by regions of the ocean to avoid autocorrelation in the data. Specifically, we designate the Atlantic ocean as test data, the eastern basin of the Pacific ocean as validation, and the remainder of the ocean as training. This is illustrated in Figure \ref{fig: train val test areas}.

\begin{figure}[ht]
    \centering
    \includegraphics[width=\textwidth]{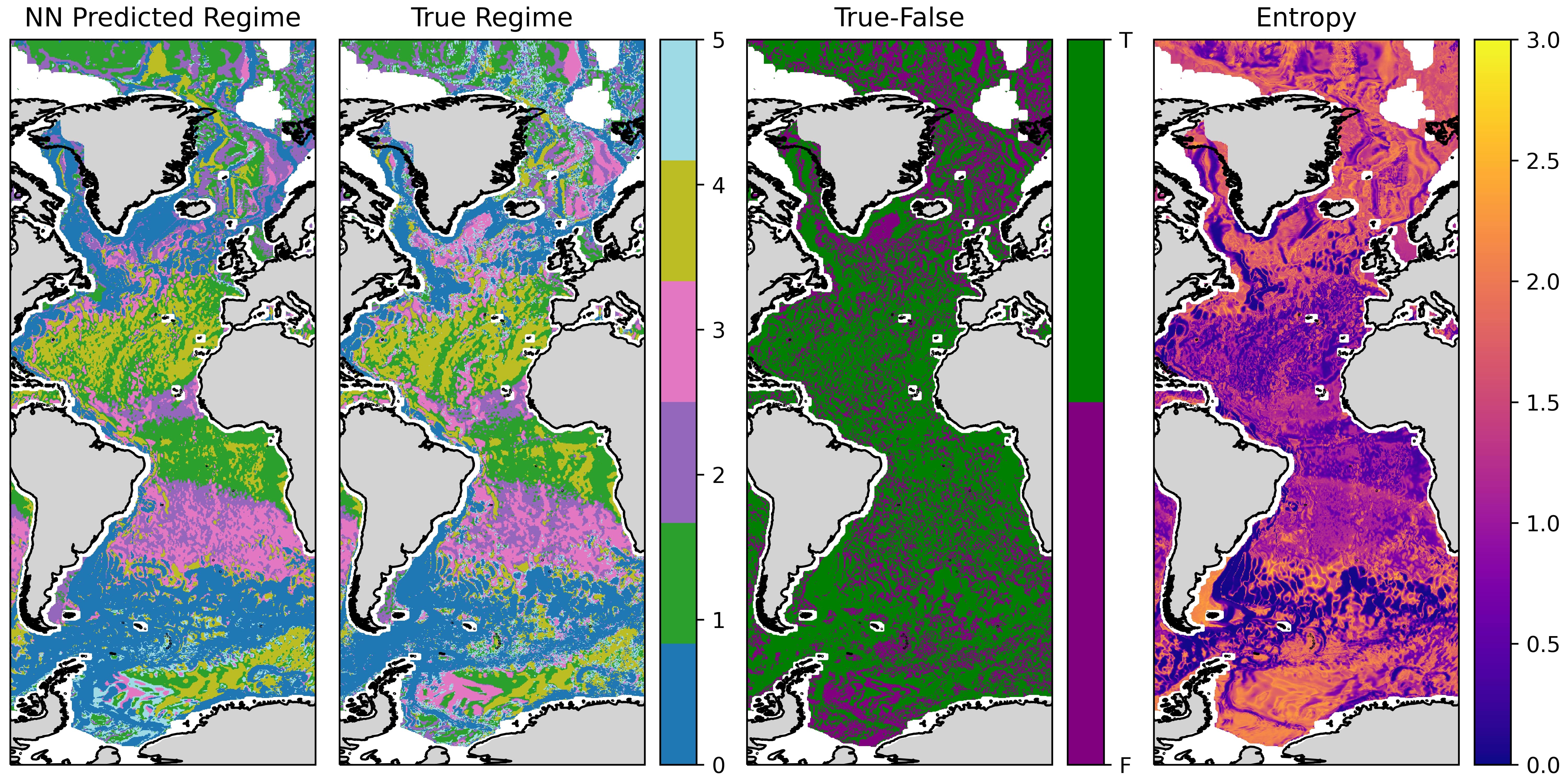}
    \caption{The NN ensemble's mean dynamical regime predictions for the test region and the true regimes. The true-false mask shows where the ensemble's mean prediction was the correct regime and where it was not. Finally, the entropy quantifies the regime's uncertainty in its predictions over the test region.}
    \label{fig: training summary}
\end{figure}

Each NN ensemble member was trained for 100 epochs to minimize categorical cross-entropy loss using the Adam optimizer with a learning rate of $1 \times 10^{-4}$ and batch size of 32. To prevent overfitting, we enforce early stopping with a patience of 5 epochs. The final validation accuracy of ensemble members ranged from 65-70\%, and the majority early stopped before 25 epochs. Figure \ref{fig: training summary} shows a visual summary of the training results. Notice that the entropy is higher in regions where THOR's deep ensemble makes more incorrect predictions, particularly in the Weddell Sea and south of Greenland. Separating grid cell entropy by correct and incorrect predictions reveals that, indeed, the average entropy for incorrect predictions is greater than that of correct predictions. This indicates that the ensemble is making well-calibrated measurements of uncertainty and that it is not overstating its own confidence in its predictions.

\begin{figure}[ht]
    \centering
    \includegraphics[width=\textwidth]{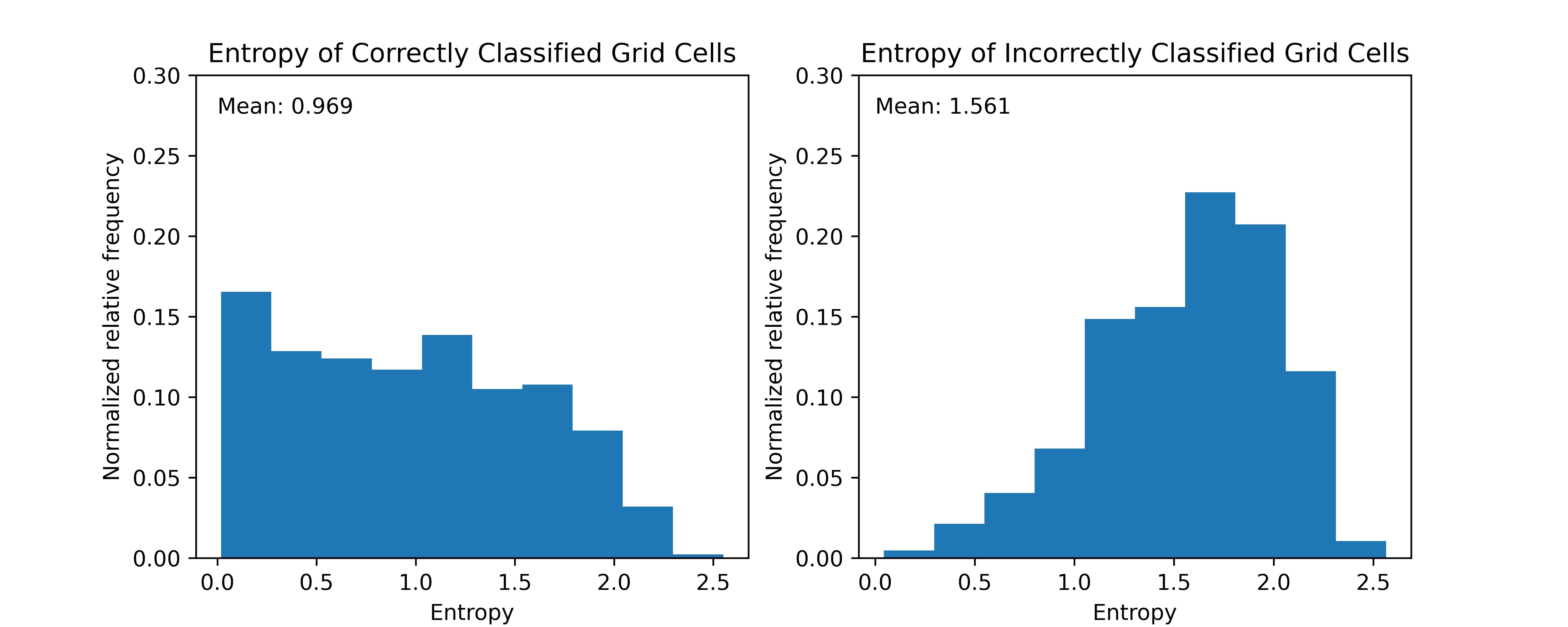}
    \caption{Entropy distributions for the correctly classified grid cells (left) and incorrectly classified grid cells (right).}
    \label{fig: entropy distributions}
\end{figure}

\subsection{XAI for the Southern Ocean}\label{appendix: xai}
\begin{figure}
    \centering
    \includegraphics[width=\textwidth]{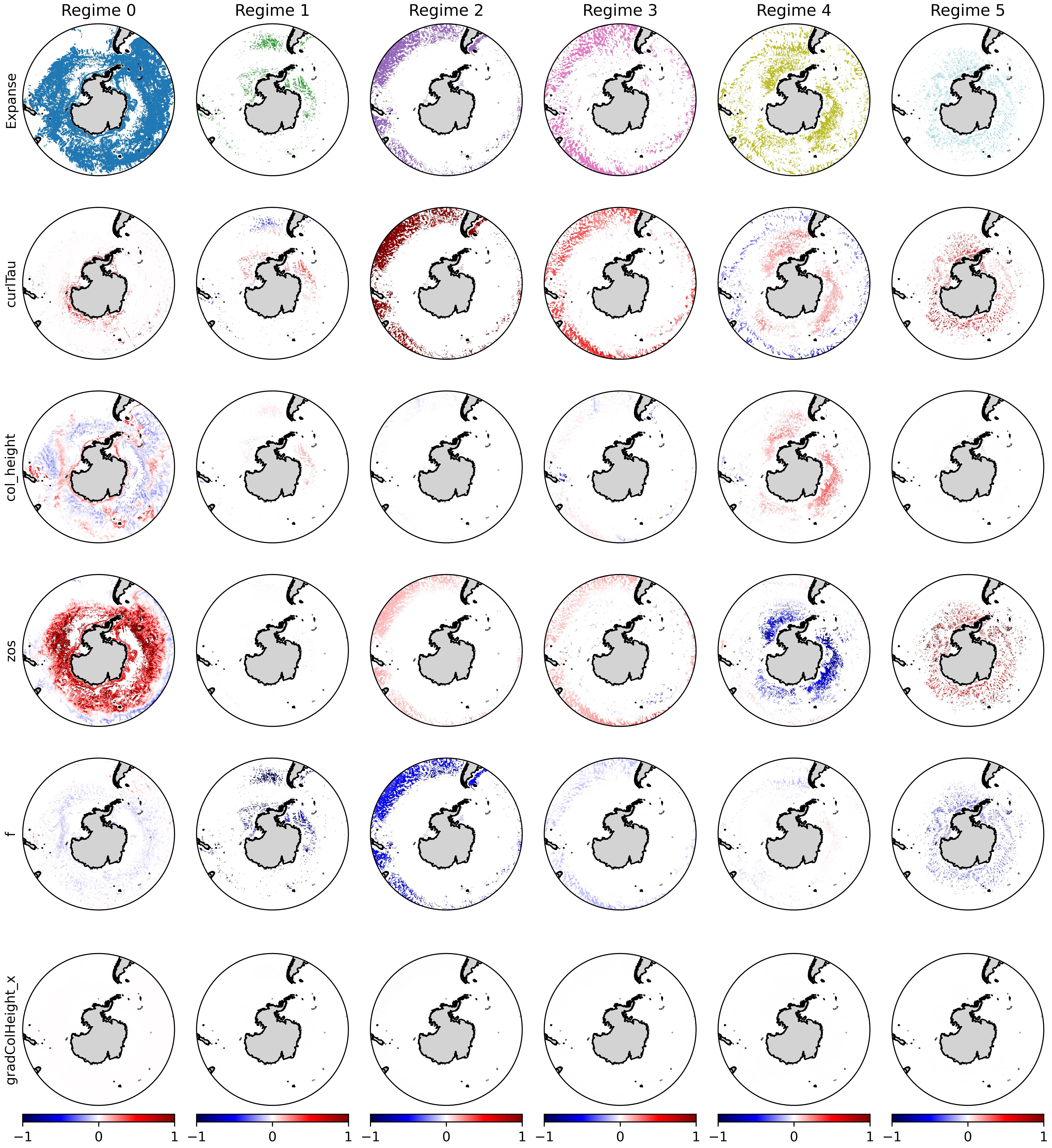}
    \caption{SSP585 scenario LRP values in the Southern Ocean for five of the NN inputs: wind stress curl, bathymetry, sea surface height, Coriolis parameter, and $x$-gradient of bathymetry. The first row shows the areas where THOR's NN ensemble predicted each regime, and each subsequent row shows the relevance of an input for predicting each regime.}
    \label{fig: lrp ssp585 first half}
\end{figure}

\begin{figure}
    \centering
    \includegraphics[width=\textwidth]{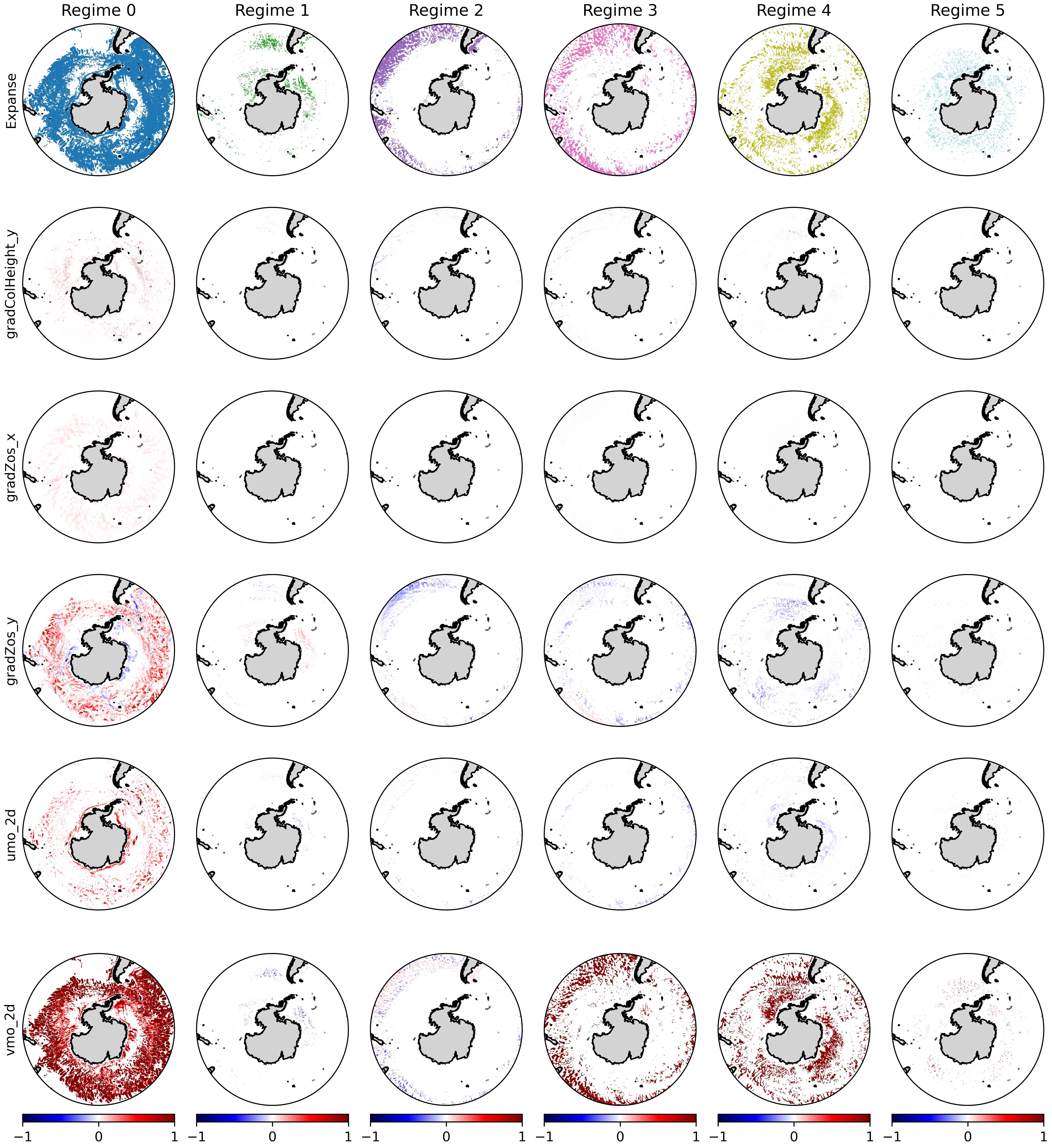}
    \caption{SSP585 scenario LRP values in the Southern Ocean for five of the NN inputs: $y$-gradient of bathymetry, $x$- and $y$-gradients of sea surface height, and depth-summed lateral and meridional mass transport. The first row shows the areas where THOR's NN ensemble predicted each regime, and each subsequent row shows the relevance of an input for predicting each regime.}
    \label{fig: lrp ssp585 second half}
\end{figure}

We apply two XAI methods to understand the relevance of each input field in the ensemble's predictions. The first is layer-wise relevance propagation (LRP), which is specific to neural network models and uses the magnitudes of internal weights as a proxy for relevance. These relevance values are backpropogated through the network to determine the importance of each input. The relevance for each input is a value between -1 and 1, inclusive. Positive values indicate that the input was actively helpful for the NN in making its prediction, negative values indicate that the input was actively unhelpful, and 0 represents neutral relevance. See \cite{montavon2019layer} for more details. We apply LRP to each of the 50 NNs of THOR's ensemble. The LRP values whose sign is consistent between the 25th and 75th percentiles of the NN ensemble in the Southern Ocean for the SSP585 (aggressive anthropogenic warming) scenario are shown in Figures \ref{fig: lrp ssp585 first half} and \ref{fig: lrp ssp585 second half}. See \cite{clare2022explainable} for more information on applying LRP to ensembles of NNs.

\begin{figure}
    \centering
    \includegraphics[width=\textwidth]{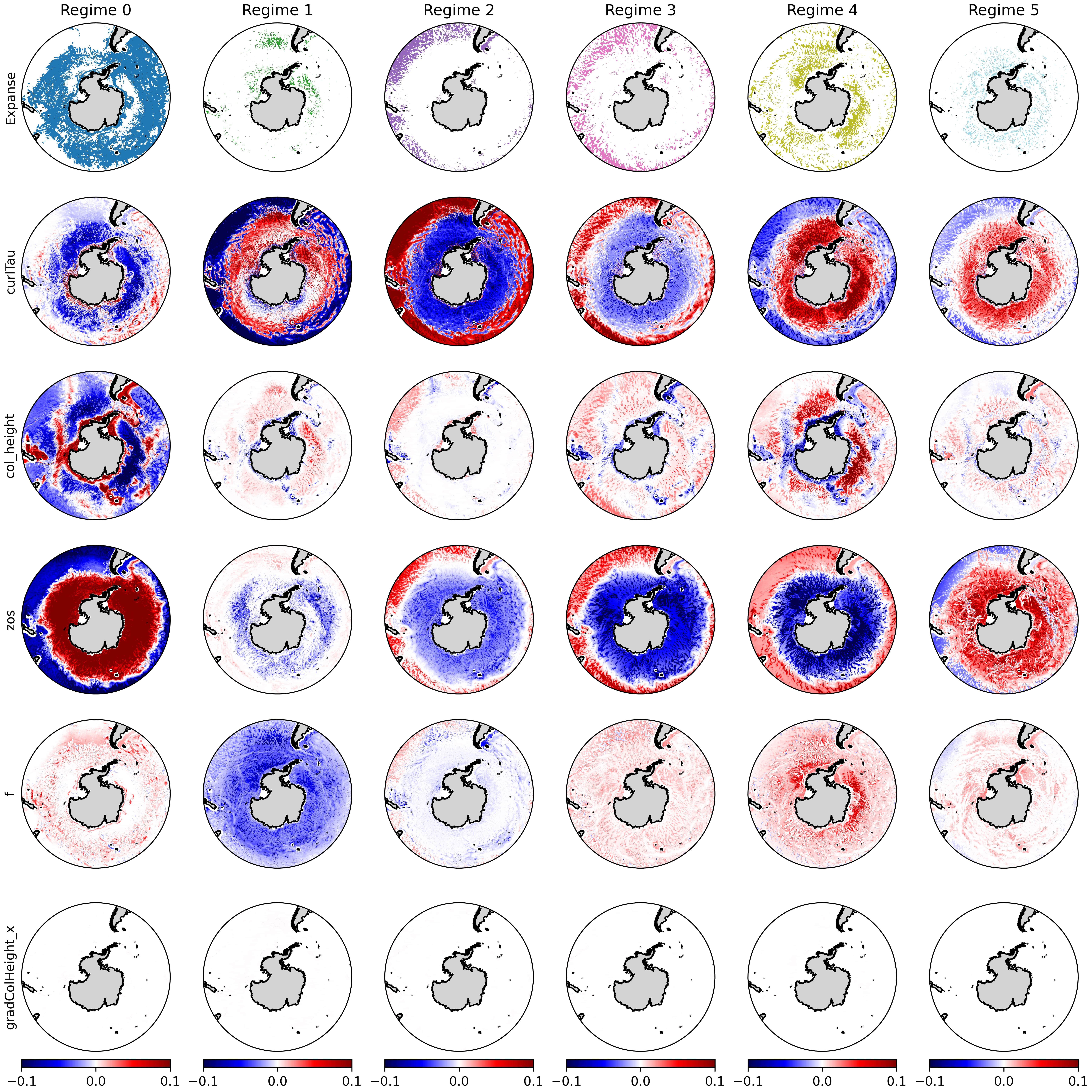}
    \caption{SSP585 scenario SHAP values in the Southern Ocean for five of the NN inputs: wind stress curl, bathymetry, sea surface height, Coriolis parameter, and $x$-gradient of bathymetry. The first row shows the areas where THOR's NN ensemble predicted each regime, and each subsequent row shows the relevance of an input for predicting each regime.}
    \label{fig: shap ssp585 first half}
\end{figure}

\begin{figure}
    \centering
    \includegraphics[width=\textwidth]{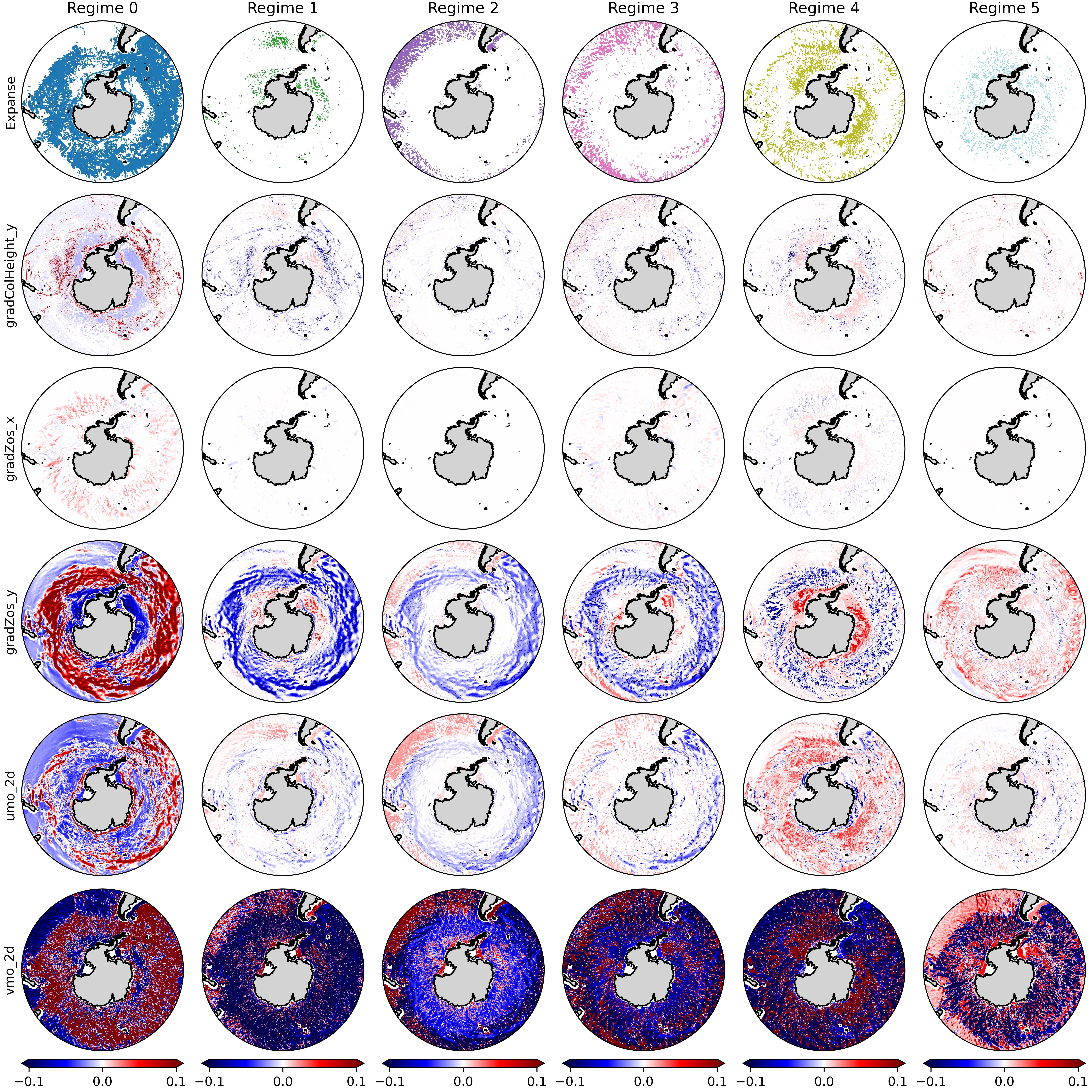}
    \caption{SSP585 scenario SHAP values in the Southern Ocean for five of the NN inputs: $y$-gradient of bathymetry, $x$- and $y$-gradients of sea surface height, and depth-summed lateral and meridional mass transport. The first row shows the areas where THOR's NN ensemble predicted each regime, and each subsequent row shows the relevance of an input for predicting each regime.}
    \label{fig: shap ssp585 second half}
\end{figure}

The second XAI method we apply is Shapley Additative exPlanations (SHAP), which is a model agnostic method for determining the relevance of each input. SHAP is a type of occlusion analysis which determines the output effect of removing/adding features from the input. For a given input and predicted output, a positive SHAP value indicates that the input increases the probability of the model predicting the output. Similarly, a negative SHAP value indicates that the input decreases the probability, and 0 indicates no change. As with LRP, we apply SHAP to each of the 50 NNs in the ensemble. The SHAP values whose sign is consistent between the 25th and 75th percentiles of the NN ensemble in the Southern Ocean for the SSP585 scenario are shown in Figures \ref{fig: shap ssp585 first half} and \ref{fig: shap ssp585 second half}. See \cite{clare2022explainable} for more information on applying SHAP to ensembles of NNs.

\end{document}